\documentclass[12pt,preprint]{emulateapj}
\usepackage{graphicx}
\usepackage{natbib} 
\usepackage{booktabs}

\newcommand{\starquantity}{37,354}
\newcommand{\exofiledate}{October 11, 2015}
\newcommand{\catalogname}{CELESTA}
\newcommand{\IRE}{\emph{Stellar Parameter Catalog}}
\newcommand{\EDE}{EDE}
\newcommand{\RHIP}{\emph{Revised Hipparcos Catalog}}
\newcommand{\TeffUncertainty}{$\pm 100 \text{K}$}
\newcommand{\RadiusUncertainty}{$\pm 0.17 R_\odot$}
\newcommand{\sMassUncertainty}{$\pm 0.10 M_\odot$}
\shorttitle{CELESTA}
\shortauthors{Colin O. Chandler, Iain McDonald \& Stephen R. Kane}
\slugcomment{Accepted for publication in The Astronomical Journal}

\begin{document}

\title{The Catalog of Earth\textendash{}Like Exoplanet Survey TArgets (CELESTA):

A Database of Habitable Zones around Nearby Stars

}

\author{
  Colin Orion Chandler\altaffilmark{1},
  Iain McDonald\altaffilmark{2},
  Stephen R. Kane\altaffilmark{1}
}
\email{coc@mail.sfsu.edu}
\altaffiltext{1}{Department of Physics \& Astronomy, San Francisco State University, 1600 Holloway Avenue, San Francisco, CA 94132, USA}
\altaffiltext{2}{Jodrell Bank Centre for Astrophysics, Alan Turing Building, Manchester, M13 9PL, UK}


\begin{abstract}
Locating planets in circumstellar Habitable Zones is a priority for many exoplanet surveys. Space\textendash{}based and ground\textendash{}based surveys alike require robust toolsets to aid in target selection and mission planning. We present the Catalog of Earth\textendash{}Like Exoplanet Survey Targets (CELESTA), a database of Habitable Zones around 37,000 nearby stars. We calculated stellar parameters, including effective temperatures, masses, and radii, and we quantified the orbital distances and periods corresponding to the circumstellar Habitable Zones. We gauged the accuracy of our predictions by contrasting CELESTA's computed parameters to observational data. We ascertain a potential return on investment by computing the number of Habitable Zones probed for a given survey duration.  A versatile framework for extending the functionality of CELESTA into the future enables ongoing comparisons to new observations, and recalculations when updates to Habitable Zone models, stellar temperatures, or parallax data become available. We expect to upgrade and expand CELESTA using data from the \emph{Gaia} mission as the data becomes available.
\end{abstract}



\section{Introduction}
\label{intro}
Target selection is crucial to the success of future exoplanetary missions, many of which aim to find Earth\textendash{}mass planets in Habitable Zones: regions around host stars where liquid water may exist on their surfaces (Huang:1959vl; Kasting et al. 1993). This a primary goal of missions in operation such as \emph{K2} \citep{Howell:2014vu}, the repurposed \emph{Kepler} spacecraft, plus upcoming missions like the \emph{Transiting Exoplanet Survey Satellite} (TESS; \citealt{Ricker:2015ie, Sullivan:2015vr}). The Habitable Zone (HZ) is determined by the properties of the host star, and the Catalog of Earth\textendash{}Like Exoplanet Survey Targets (CELESTA) we present here computes the necessary parameters, enabling CELESTA to serve as a target selection tool to assist in mission planning and operations.

Catalogs such as the Catalog of Nearby Habitable Systems (HabCat; \citealt{Turnbull:2003em}), have concentrated on individual stellar abundances in relation to habitability considerations. Work dependent on the original \emph{Hipparcos} catalog would benefit from the improvements \cite{VanLeeuwen:2007uqb} made in \RHIP{}, further detailed in Section \ref{HIP2}. CELESTA contains \starquantity{} stars, up from 17,129 \emph{Hipparcos} stars found in HabCat \citep{Turnbull:2003em}. 
Furthermore, advances have been made in the understanding of HZ properties (e.g. \citealt{Kopparapu:2014tg}) and stellar temperatures (e.g. \citealt{McDonald:2012tx}) in the last few years. Existing works that make use of HabCat will also benefit from the upgrades provided by CELESTA, especially by the improved temperature and HZ distance calculations.

The most critical element to target selection is the understanding of the stars themselves. In this work, we derive the stellar parameters and HZ radii for stars of the \RHIP{} \citep{VanLeeuwen:2007uqb}, a re\textendash{}reduction of the original \emph{Hipparcos} catalog \citep{Perryman:1997vj}, the largest sample of accurate stellar parallactic distances to date. The \RHIP{} contains 117,955 bright, nearby stars, along with parallax data from which we can calculate distances. 
\cite{McDonald:2012tx} built upon the \RHIP{} and created a catalog of 107,619 stars that included stellar temperatures and luminosities. Kopparapu et al (2013, 2014), furthering work from \cite{Kasting:1993wp}, put forth a set of coefficients for stellar parameters to be used in determining the boundaries of HZs. The Exoplanet Data Explorer \citep{Wright:2011vs} providesd a comparison sample against which to test our catalog. Looking to the future, the \emph{Gaia} mission \citep{deBruijne:2015wb} will allow for upgrades to CELESTA, simultaneously increasing the number of stars and the precision of the calculations contained in CELESTA.

This analysis results in a catalog of \starquantity{} stars with their expected HZs, using data from multiple sources. In Section \ref{hzs} we provide a brief synopsis of the data and terminology used in the analysis of habitable zones around exoplanets. In Section \ref{CatalogCreation} we detail the creation of CELESTA, including describing the data sources utilized and our methods for selecting stars to be included. A sample from CELESTA\footnote{See the online publication for the complete dataset.} is contained in Section \ref{catalog}. Section \ref{Applications} covers example applications of CELESTA.


\section{Background: Habitable Zones}
\label{hzs}
A \emph{Habitable Zone} (HZ), also known as a Goldilocks Zone or a Comfort Zone, can be categorized into four classes \citep{Forget:2012gx}. Class I, the only class considered by this paper, covers Earth\textendash{}like planets residing in a shell around one or more host stars where a planet or other body could potentially harbor liquid water on its surface \citep{Huang:1959vl,Kasting:1993wp} and receive sunlight. Class II zones are regions where planets no longer possess liquid water on their surfaces, e.g. Mars or Venus. Class III describes worlds with liquid water below the surface but in direct contact with a silicate core e.g. Europa, and Class IV describes planets with liquid water trapped between layers of ice, e.g. Io.  For linguistic convenience we shall refer to the host(s) as a \emph{star}, and the objects orbiting these stars as \emph{planets}. Planets on eccentric orbits may pass in and out of the HZ \citep{Kane:2012di}. Here we only consider the boundaries of the HZ, without exploring the orbits of the planets themselves.

\begin{figure}[h]
\begin{center}
\includegraphics[width=.9\linewidth]{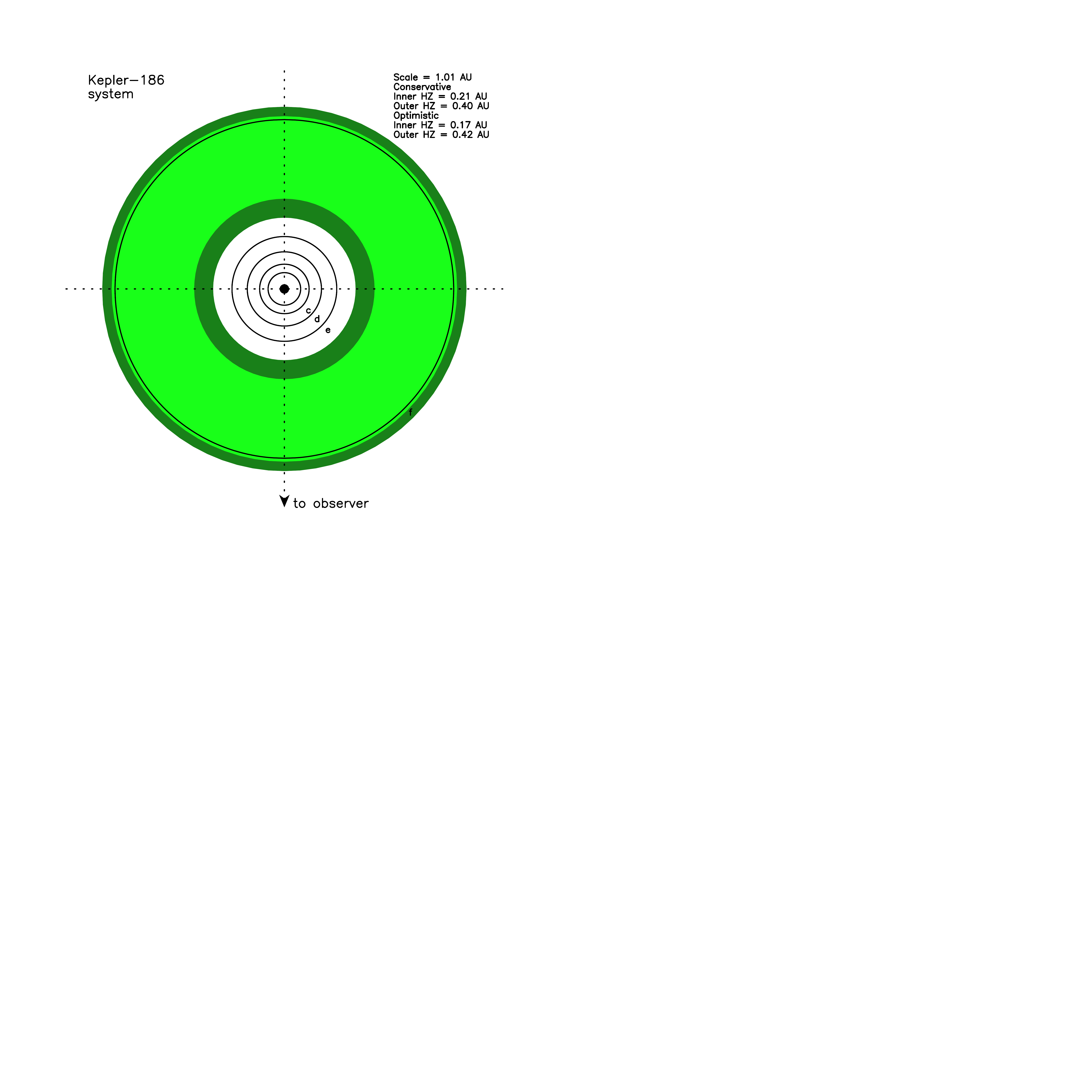}
\caption{HZ regions around a single host star, Kepler\textendash{}186, in this example provided by The Habitable Zone Gallery (http://hzgallery.org; \citealt{Kane:2012co}) and based on the work of \cite{Quintana:2014cc}; the Conservative Habitable Zone (dark green) runs between the inner Runaway Greenhouse and outer Maximum Greenhouse borders. The Optimistic Habitable Zone (light green) lies between the inner boundary, Recent Venus, and the outer Early Mars border.}
\label{HZRegions}
\end{center}
\end{figure}

The HZ is oft considered from two different perspectives: the Optimistic Habitable Zone (OHZ) and Conservative Habitable Zone (CHZ). The OHZ and CHZ each contain an inner boundary ($\text{H}_{2}\text{O}$ dominated atmosphere) and outer boundary ($\text{CO}_{2}$ dominated atmosphere): the OHZ is bound on the inside (closest to the star) by the Recent Venus boundary and on the outside (furthest from the star) by the Early Mars limit; the CHZ lies between the inner Runaway Greenhouse border and the outer Maximum Greenhouse cutoff. Figure \ref{HZRegions}, courtesy of the Habitable Zone Gallery (http://hzgallery.org; \citealt{Kane:2012co}), portrays these regions graphically. \cite{Kopparapu:2013wba,Kopparapu:2014tg} provide details of these divisions. ``Earth\textendash{}like'' exoplanets, for the context of this paper, refers to planets that fall within the terrestrial regime of $R_{\text{planet}} \leq 1.5 R_\oplus$ as described by \cite{Rogers:2015jn}.

\section{Catalog Creation}
\label{CatalogCreation}

\subsection{Input Data Sources}
\label{sources}
The first step in creating CELESTA was assembling the input data. The following is a brief description of each source and how that source pertains to CELESTA.

\subsubsection{The \emph{Revised Hipparcos Catalog}}
\label{HIP2}
The \RHIP{} \citep{VanLeeuwen:2007uqb} is a stellar catalog based on the original \emph{Hipparcos} mission \citep{Perryman:1997vj} dataset. \emph{Hipparcos}, launched in 1989, recorded with great precision the parallax of nearby stars, ultimately leading to a database of 118,218 stars. The \RHIP{} was a refinement of this original catalog, applying newer methods in order to reduce error in areas such as parallax, resulting in a final catalog containing 117,955 stars.

\subsubsection{Stellar Parameter Catalog}
\label{IRE}
\citet{McDonald:2012tx} calculated effective temperatures and luminosities for the \emph{Hipparcos} stars. In brief, their calculation compares the {\sc bt-settl} stellar model atmospheres \citep{Allard:2003waa} to observed optical and infrared broadband photometry and derives effective temperatures on the basis of a $\chi^2$ statistic, scaling the model atmospheres in flux to provide stellar luminosities. In this work, we repeat the data reduction from that paper. This new reduction uses an identical pipeline but with an improved interpolation between stellar model atmospheres to remove striping effects in the Hertzsprung--Russell diagram.

\begin{figure}
\begin{center}
\includegraphics[width=1.0\linewidth]{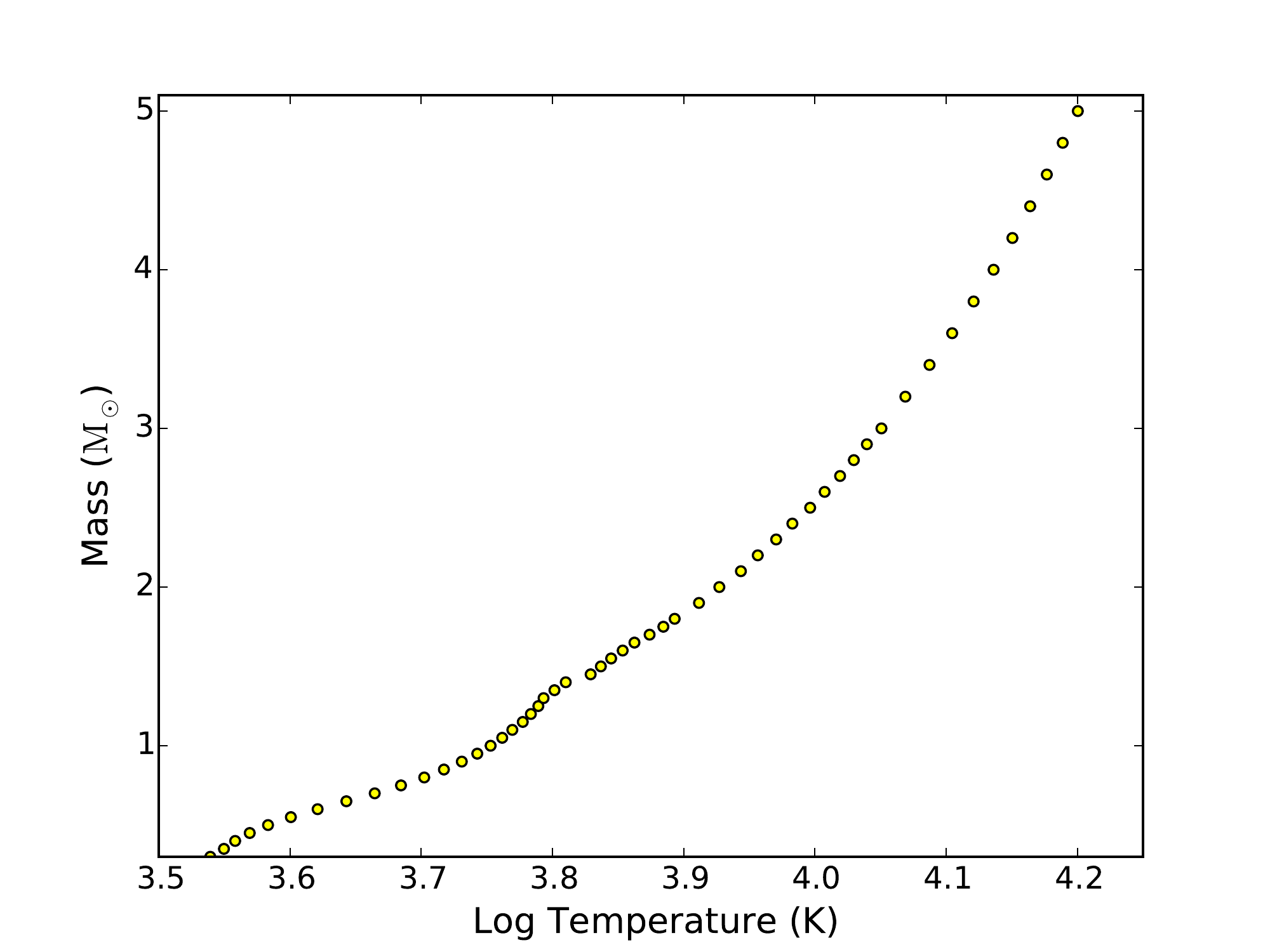}
\caption{The conversion of main\textendash{}sequence effective temperature to stellar mass, based on the stellar evolution models of \cite{Dotter:2008ti}.}
\label{MassInterpolation}
\end{center}
\end{figure}

First, a blackbody is fit to derive an approximate effective temperature and luminosity. On this basis, the star is classified as a giant or main\textendash{}sequence star. Giant stars are assumed to be 1 M$_\odot$\footnote{The initial mass function (e.g. \citealt{Kroupa:2001vq}) and the declining star formation in the Solar Neighborhood since the birth of the Milky Way (e.g. \citealt{Rowell:2013cr,Haywood:U-t7X7d8}) strongly bias giant stars to be of low mass. The oldest thin-disc stars should have $M \approx 0.90$--0.96 M$_\odot$ (e.g. \citealt{Bressan:2014iz}). A few percent of stars may have lower masses (e.g. metal-poor thick disc or halo stars, or those which have suffered considerable mass loss at the RGB tip; \citealt{Gilmore:1983ip,McDonald:2015fy}). The average giant star mass should therefore be $\sim$1 M$_\odot$.} For the SED-fitting process, stellar mass only sets the surface gravity, therefore only has a second-order impact on the derived parameters. We do not compute habitable zones for giant stars in this work. The mass-temperature relation CELESTA employed is illustrated in Figure \ref{MassInterpolation}.

\begin{table*}
\caption{Sample data from the \IRE{}}
\label{IRE2table}
\begin{center}
\begin{tabular}{c|c|c|c|c|c|c|c|c}
\toprule
\midrule
HIP & {RA (deg)} & {Dec (deg)} & $T_{\text{eff}}$ (K)\footnote{Uncertainty in $T_{\text{eff}}$ was determined to be \mbox{\TeffUncertainty{}} by comparing the \IRE{} predictions to observations contained in \EDE{}. Significant figures were preserved from the \IRE{} input file; see the text (Sections 3 and 4) for comments on additional uncertainties.} & {$L_\star (L_\odot)$} & {Blue (nm)\footnote{\label{note1}The bluest and reddest wavelengths for which data existed in the matched photometric catalog.}} & {Red (nm) \footnote{See footnote \ref{note1}.}} & {$d$ (pc)} & {$\delta d/d$}\\
\hline
1 & 0.00091185 & 1.08901332 & 6388 & 8.73 & 354 & 2200 & 220 & 0.29\\
2 & 0.00379738 & -19.49883738 & 4506 & 0.613 & 354 & 623 & 48 & 0.05\\
3 & 0.00500794 & 38.85928608 & 8968 & 375 & 420 & 8610 & 442 & 0.16\\
4 & 0.00838188 & -51.89354611 & 7005 & 8.45 & 420 & 2200 & 134 & 0.1\\
5 & 0.00996502 & -40.59122445 & 5064 & 27.4 & 420 & 8610 & 258 & 0.24\\
8 & 0.02729175 & 25.8864746 & 2992 & 743 & 354 & 25000 & 201 & 0.37\\
9 & 0.03534194 & 36.58593769 & 4500 & 81.2 & 354 & 8610 & 420 & 0.39\\
10 & 0.03625296 & -50.86707363 & 6475 & 2.56 & 420 & 2200 & 92.4 & 0.09\\
11 & 0.03729667 & 46.94000168 & 8188 & 51.8 & 420 & 2200 & 239 & 0.14\\
{\vdots} & {\vdots} & {\vdots} & {\vdots} & {\vdots} & {\vdots} & {\vdots} & {\vdots} & {\vdots}\\
113541 & 344.9415731 & -31.12644206 & 3658 & 2550 & 420 & 18390 & 909 & 0.75\\
113542 & 344.9417062 & 25.64387254 & 6833 & 2.24 & 354 & 2200 & 94.9 & 0.08\\
\bottomrule
\end{tabular}

\end{center}
\end{table*}

For this paper, the original catalog from \citet{McDonald:2012tx} was re\textendash{}reduced using a slightly updated version of the spectral energy distribution pipeline, in order to correct for an interpolation artifact when temperatures were close to those of the stellar atmosphere models (every 100 K). With this exception, the \IRE{} presented here is functionally identical to that described in \citet{McDonald:2012tx}. Identical cuts have been made to remove stars where accurate parameters could not be determined, either because the source could not be well\textendash{}modeled by a single stellar spectrum or because there was insufficient good\textendash{}quality data to model, leaving 103,663 out of the original 117,955 stars with stellar parameters. At this stage, we retained stars where parallaxes were highly uncertain. Fractional error in parallax was determined by the ratio of error in parallax to the measured parallax, both values originating from the \RHIP{}. The $3700+$ stars with negative parallaxes were excluded from CELESTA. A sample from the \IRE{} is included in Table \ref{IRE2table}.

\subsubsection{Habitable Zone Boundary Parameters}
\cite{Kopparapu:2014tg} provide a means to calculate the effective solar flux, $S_{\text{eff}}$, and HZ boundary distances and widths for each star. The parametric equations provided by Equation 4 of \cite{Kopparapu:2014tg} for calculating the effective solar flux were:
\begin{equation}
S_{\text{eff}}=S_{\text{eff}_\odot}+a T+b T^2+c T^3+d T^4
\label{seffequation}
\end{equation}
where $T=T_{\text{eff}}-5780 K$, and the coefficients for each of the four were provided in the \cite{Kopparapu:2014tg} paper.

In each of the four cases, the corresponding distance ($d$) of the HZ boundary can be calculated using Equation 5 of \cite{Kopparapu:2014tg}:
\begin{equation}
d=\sqrt{\frac{L/L_{\odot}}{S_{\text{eff}}}} \text{AU}
\label{distanceequation}
\end{equation}
with $L/L_\odot$ being the ratio of the star's luminosity compared to the luminosity of our Sun.

Note that \cite{Kopparapu:2014tg} stipulate that the coefficients they provide are valid for $0.1 M_\oplus - 5 M_\oplus$ planets only. They also warn that if an exoplanet is tidally locked to its host star, the Inner HZ could be expanded inward, as described by \cite{Yang:2013gl}. These scenarios, along with the Class II \textendash{} Class IV HZs described in Section \ref{hzs}, are outside the scope of this paper.


\subsection{Star Selection}
\label{starselection}

\begin{figure}
\begin{center}
\includegraphics[width=1.1\linewidth]{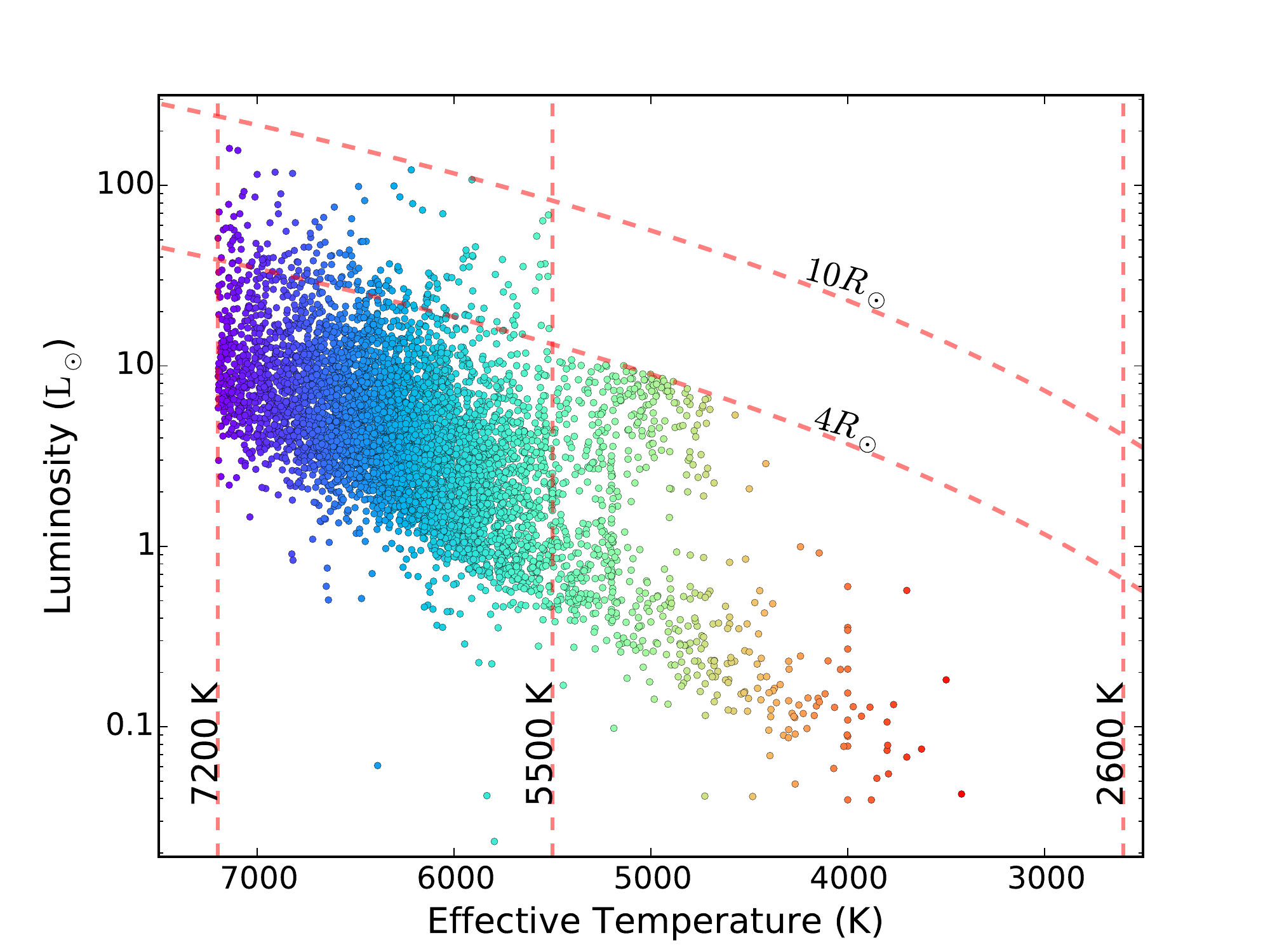}
\caption{H\textendash{}R Diagram of the stars contained in CELESTA after cuts were applied; every fifth star was chosen for plotting. Note the cut\textendash{}offs applied to remove giant stars with radii above 10 $R_\odot$ ($>$5500 K) and 4 $R_\odot$ ($<$5500 K). Stars outside the range of 2,600 K \textendash{} 7,200 K were also rejected.}
\label{HRD}
\end{center}
\end{figure}

The next step was selecting appropriate stars for the construction of CELESTA . The \IRE{} of 103,663 stars included many stars that were not suitable to our purposes, especially stars off the main\textendash{}sequence (MS) branch, e.g. giants. The following selection criteria were used in creating the final catalog. Unless indicated otherwise, these selection criteria are mutually exclusive, and stars may have been rejected by one or more of these criteria.

\cite{Kopparapu:2014tg} stipulate that the HZ coefficients they provide are only valid for stars between 2,600 K and 7,200 K. Figure \ref{HRD} contains dashed vertical lines at each of these temperatures. 41 stars below 2,600 K and 23,926 stars above 7,200 K were ineligible for inclusion in CELESTA. Since the work of \cite{Kopparapu:2014tg} only applied to MS stars, we filtered stars based upon their radii. The radius of the star, $R$, can be found using the familiar Stefan\textendash{}Boltzmann equation:

\begin{equation}
L = A \sigma T^4
\end{equation}
\label{sbequation}
where $L$ is the stellar luminosity, $A$ is the surface area, $\sigma$ is the Stefan\textendash{}Boltzmann constant, and \textit{T} is the stellar effective temperature. Using the surface area of a sphere, $4\pi R^2$, and solving for the radius, \textit{R}: 

\begin{equation}
R=\sqrt{\frac{L}{4 \pi \sigma T^4}}
\label{radius}
\end{equation}

Giants were considered to be stars falling in either of the following two categories: stars with $R > 10R_\odot$, and stars with both $R > 4 R_\odot$ and $T < 5500 K$. The former excluded 34,169 stars and the latter 10,082 stars. Figure \ref{HRD} includes dashed lines delineating these boundaries. We also found ineligible 31 stars with $R < 0.1 R_\odot$.

At present, no cuts were made based on uncertainty in parallax, though stars with $>$30\% uncertainty in parallax were removed for the analysis that follows. We discuss error in parallax further in Section \ref{errors}.

\subsection{Orbital Periods}
\label{OrbitalPeriods}

Orbital periods were derived from Kepler's third law, namely:
\begin{equation}
P = \frac{1}{86\,400} \sqrt{\frac{4 \pi^2}{G(M_{\ast}/M_{\odot})}\left(\frac{r}{\rm{1AU}}\right)} \  \rm{days}
\label{PeriodEquation}
\end{equation}
where $r$ is the orbital radius and $M_\ast$ is the stellar mass.


\section{CELESTA: The Catalog}
\label{catalog}

\newcommand{\magfoot}{Significant figures were carried from the source data where applicable.}
\newcommand{\distfoot}{The online edition also contains a sexigecimal right ascension \& declination field, omitted here due to the availability of printed space.}
\newcommand{\tempfoot}{Uncertainty in $T_{\text{eff}}$ was determined to be \TeffUncertainty{}.}
\newcommand{\massfoot}{Uncertainty in $M_\odot$ was determined to be \sMassUncertainty{}.}
\newcommand{\radfoot}{Uncertainty in $R_\odot$ was determined to be \RadiusUncertainty{}.}
\newcommand{\edesignote}{Signifigant figures carried from the \EDE{}.}
\begin{table*}
\caption{Sample stars from CELESTA.}
\label{HZTable}
\begin{center}
\begin{tabular}{lrrrrrrrr}
\hline
 \emph{Hipparcos} Number                            		& 1640     & 1666      & 1931      & 3502       & 6379      & 6511      & 7562      & 7599       \\
 Distance (pc)\footnote{\distfoot}                       		&   79     &  108      &  100      &   77       &   17      &   53      &   99      &   39       \\
 $\delta$ Distance (pc)                      				&    4     &    9      &   10      &    6       &    0.13   &    1.7    &   10      &    0.6     \\
 Stellar Temperature (K)\footnote{\tempfoot}             	& 5018     & 5975      & 6013      & 5426       & 5213      & 6223      & 6538      & 6013       \\
 Magnitude\footnote{\magfoot}                      		&    7.687 &    8.2861 &    9.1385 &    8.8454  &    7.3101 &    7.8192 &    9.3927 &    7.4622  \\\
 Luminosity $(L_\odot)$                      				&    6.7   &    4.8    &    2.23   &    1.7     &    0.39   &    1.9    &    1.6    &    1.5     \\
 $\delta$ Luminosity $(L_\odot)$                    		&    0.9   &    0.9    &    0.5    &    0.3     &    0.03   &    0.17   &    0.3    &    0.11    \\
 Stellar Mass $(M_\odot)$\footnote{\massfoot}         	&    0.79  &    1.14   &    1.16   &    0.915   &    0.85   &    1.3    &    1.4    &    1.16    \\
 Stellar Radius $(R_\odot)$\footnote{\radfoot}      	&    3.4   &    2.05   &    1.4    &    1.5     &    0.77   &    1.19   &    0.98   &    1.14    \\
 Recent Venus $S_{\text{eff}}$                			&    1.6   &    1.8    &    1.8    &    1.7     &    1.7    &    1.9    &    1.9    &    1.8     \\
 $\delta$ Recent Venus $S_{\text{eff}}$               	&    0.2   &    0.3    &    0.4    &    0.3     &    0.23   &    0.4    &    0.4    &    0.4     \\
 Runaway Greenhouse $S_{\text{eff}}$          		&    1     &    1.13   &    1.14   &    1.06    &    1      &    1.17   &    1.21   &    1.14    \\
 $\delta$ Runaway Greenhouse $S_{\text{eff}}$      	&    0.13  &    0.21   &    0.22   &    0.16    &    0.14   &    0.24   &    0.3    &    0.22    \\
 Maximum Greenhouse $S_{\text{eff}}$          		&    0.311 &    0.37   &    0.37   &    0.33    &    0.322  &    0.38   &    0.4    &    0.37    \\
 $\delta$ Maximum Greenhouse $S_{\text{eff}}$      	&    0.04  &    0.07   &    0.07   &    0.05    &    0.05   &    0.08   &    0.09   &    0.07    \\
 Early Mars $S_{\text{eff}}$                  			&    0.28  &    0.33   &    0.33   &    0.3     &    0.29   &    0.34   &    0.36   &    0.33    \\
 $\delta$ Early Mars $S_{\text{eff}}$                 		&    0.04  &    0.06   &    0.07   &    0.05    &    0.04   &    0.07   &    0.08   &    0.07    \\
 Recent Venus Distance (AU)            				&    2     &    1.6    &    1.1    &    0.99    &    0.48   &    1      &    0.9    &    0.913   \\
 $\delta$ Recent Venus Distance (AU)           		&    0.07  &    0.09   &    0.05   &    0.03    &    0.003  &    0.03   &    0.04   &    0.017   \\
 Runaway Greenhouse Distance (AU)       		&    2.6   &    2.07   &    1.4    &    1.25    &    0.614  &    1.3    &    1.14   &    1.16    \\
 $\delta$ Runaway Greenhouse Distance (AU)      	&    0.11  &    0.14   &    0.08   &    0.04    &    0.005  &    0.04   &    0.07   &    0.03    \\
 Maximum Greenhouse Distance (AU)      		&    4.7   &    3.6    &    2.5    &    2.22    &    1.1    &    2.24   &    2      &    2       \\
 $\delta$ Maximum Greenhouse Distance (AU)     	&    0.4   &    0.5    &    0.24   &    0.14    &    0.017  &    0.13   &    0.2    &    0.09    \\
 Early Mars Distance (AU)              				&    4.9   &    3.8    &    2.6    &    2.3     &    1.16   &    2.4    &    2.08   &    2.14    \\
 $\delta$ Early Mars Distance (AU)             		&    0.4   &    0.5    &    0.3    &    0.16    &    0.019  &    0.14   &    0.22   &    0.099   \\
 Conservative HZ Width (AU)        				&    2.1   &    1.6    &    1.1    &    0.98    &    0.5    &    0.96   &    0.8    &    0.9     \\
 Optimistic HZ Width (AU)(AU)          				&    3     &    2.2    &    1.5    &    1.4     &    0.7    &    1.3    &    1.2    &    1.2     \\
 Recent Venus Period (days)        				& 1190     &  712      &  390      &  370       &  130      &  330      &  260      &  300       \\
 $\delta$ Recent Venus Period (days)       		&  100     &   70      &   30      &   30       &    8      &   20      &   20      &   20       \\
 Runaway Greenhouse Period (days)  			& 1700     & 1000      &  560      &  530       &  190      &  460      &  370      &  421       \\
 $\delta$ Runaway Greenhouse Period (days) 	&  200     &  100      &   50      &   40       &   10      &   30      &   30      &   20       \\
 Maximum Greenhouse Period (days)  			& 4110     & 2400      & 1300      & 1300       &  460      & 1070      &  850      &  980       \\
 $\delta$ Maximum Greenhouse Period (days) 	&  600     &  500      &  200      &  100       &   30      &  100      &  100      &   80       \\
 Early Mars Period (days)          					& 4500     & 2600      & 1400      & 1400       &  500      & 1160      &  922      & 1060       \\
 $\delta$ Early Mars Period (days)         			&  700     &  500      &  200      &  200       &   30      &  100      &  100      &   90       \\
 Parallax                            						&   12.6   &    9.22   &    9.75   &   12.95    &   59.49   &   19.03   &   10.06   &   25.63    \\
 $\delta$ Parallax                          				&    0.65  &    0.76   &    0.95   &    1.03    &    0.46   &    0.6    &    1.07   &    0.38    \\
 EDE Exoplanet Count                 				&    2     &    1      &    1      &    1       &    3      &    1      &    1      &    6       \\
 EDE Radius $(R_\odot)$\footnote{\edesignote{}}  	&    3.8   &    1.93   &    1.1465 &    1.15783 &    0.7821 &    1.0437 &    1.216  &    1.10851 \\
 EDE Stellar Temperature (K)                     			& 4757     & 6317      & 5940.77   & 5701.69    & 5177      & 6136      & 6400      & 5911       \\
 EDE Stellar Mass $(M_\odot)$                 			&    1.31  &    1.5    &    1.123  &    1.13    &    0.832  &    1.13   &    1.22   &    1.06    \\
\hline
\end{tabular}

\end{center}
\end{table*}

The final CELESTA catalog contains \starquantity{} stars, each with a set of associated attributes, e.g. estimated mass, measured distance. A sample from CELESTA can be found in Table \ref{HZTable}, and the complete database can be found online at a dedicated host\footnote{http://celesta.info} as well as the VizieR repository\footnote{http://vizier.u-strasbg.fr}.

\begin{figure*}
\begin{center}
\includegraphics[width=1\linewidth]{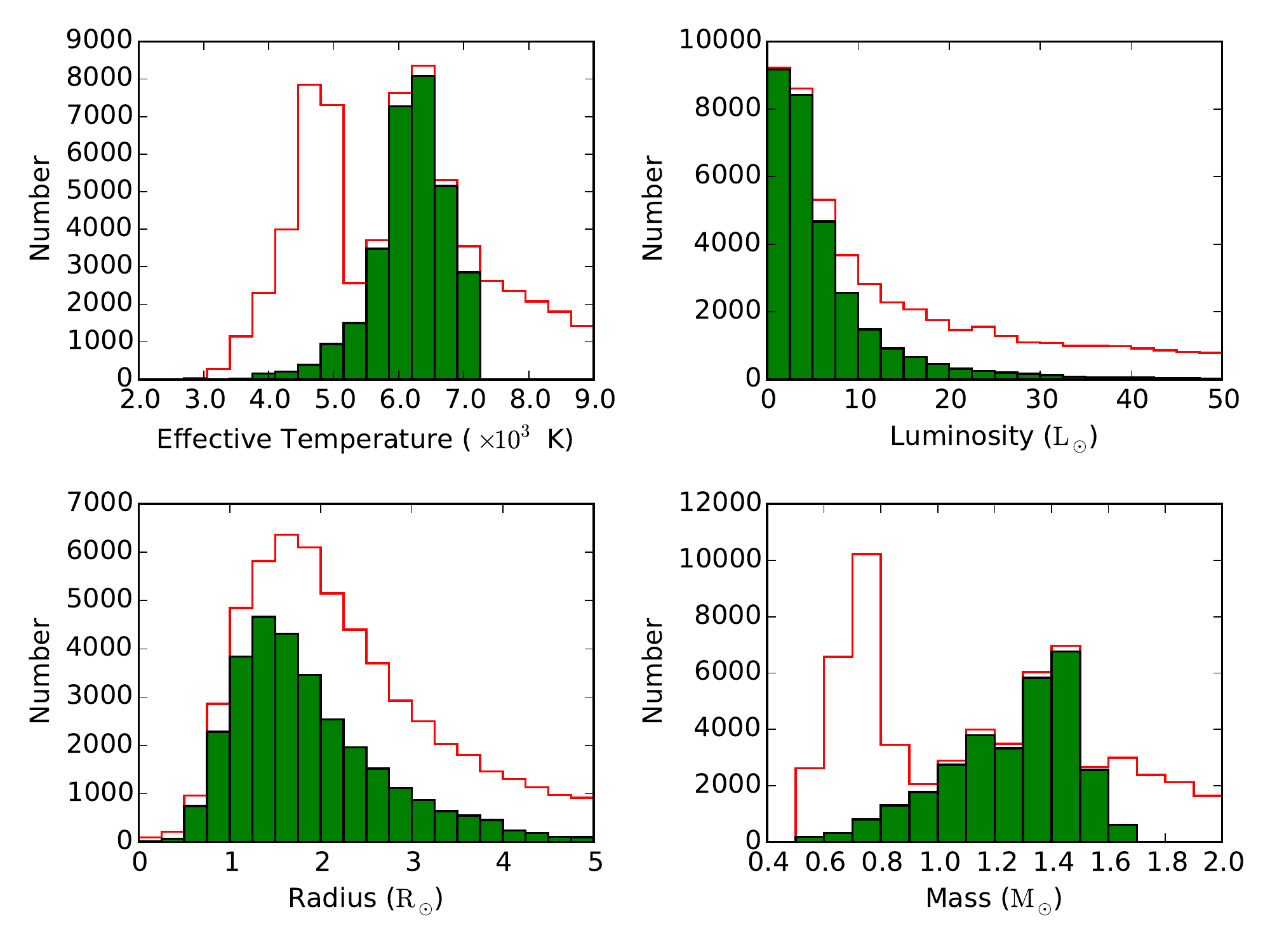}
\caption{Comparisons of stellar parameters between CELESTA and the \IRE{}. Only the most populous sections of the histogram are shown. \emph{Upper Left}: The lack of low temperature stars in CELESTA stems from the exclusion of giant stars, and the upper dropoff from the requirements of \cite{Kopparapu:2014tg}. \emph{Upper Right}: the removal of stars from CELESTA increased as the luminosity increased. \emph{Lower Left}: The cuts by radius were more pronounced after $1 R_\odot$. \emph{Lower Right}: The large number of low\textendash{}mass ($0.5 M_\odot$ \textendash{} $1.0 M_\odot$) stars excluded from CELESTA were mostly giants.}
\label{MultiHist}
\end{center}
\end{figure*}

Comparison of the stars with $<30\%$ error in parallax found in both CELESTA and the underlying \IRE{} provided some insights, as seen in Figure \ref{MultiHist}. To begin with, we see that the distribution of stars ineligible for inclusion in CELESTA (as indicated by the unfilled sections of Figure \ref{MultiHist}) varied by stellar parameter. For example, the temperature spread found in the \IRE{} was more even than with CELESTA, whose concentration was concentrated around 6,000 K; this resulted from the removal of the cooler giants, as was the drop\textendash{}off of low\textendash{}mass stars. Few stars near Solar luminosity ($<$5 L$_\odot$) were removed, and the cuts in radius (Figure \ref{MultiHist}) was fairly uniform.

\begin{figure}
\begin{center}
\includegraphics[width=1\linewidth]{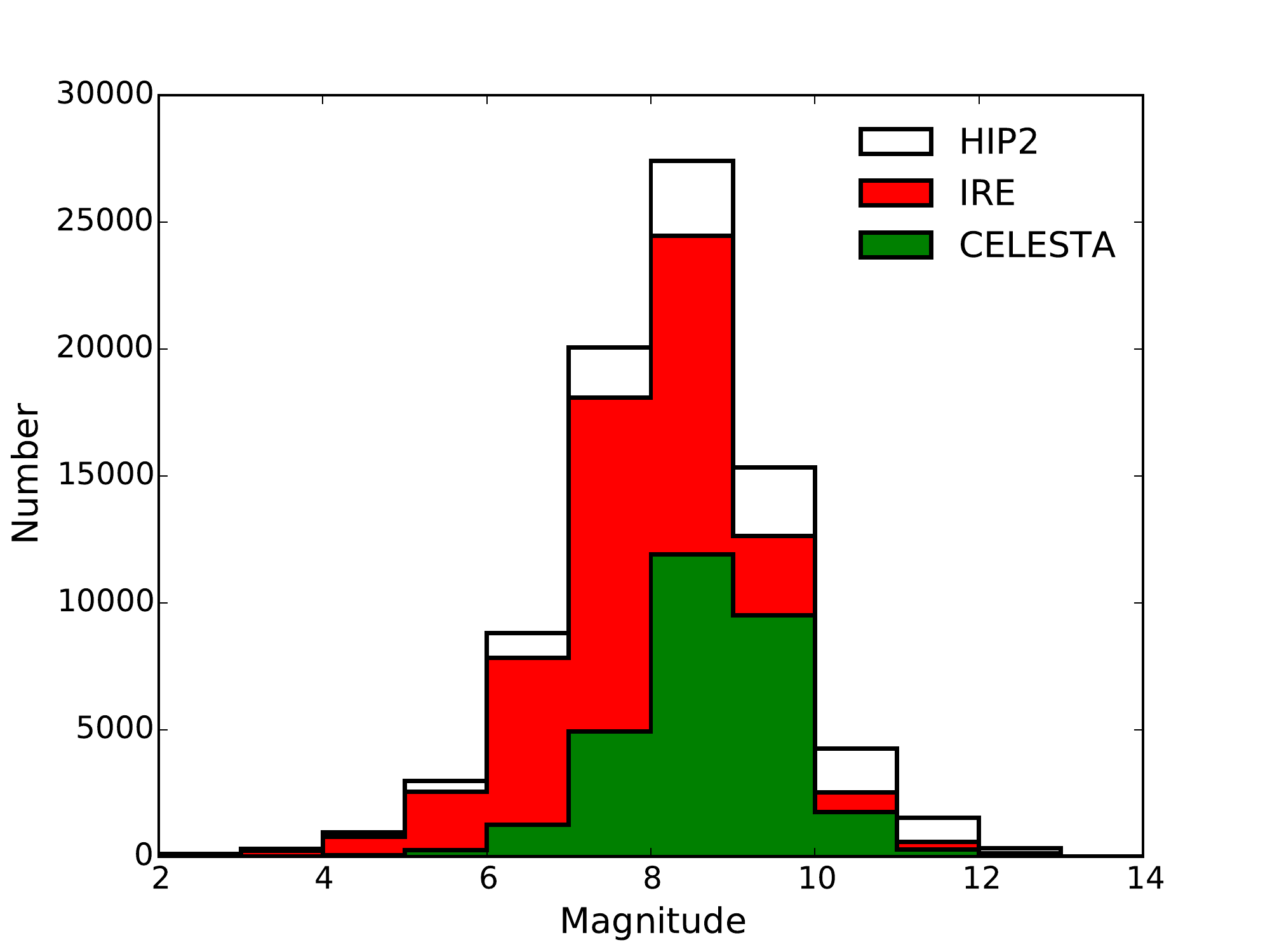}
\caption{The magnitude distribution of stars covered by CELESTA, \RHIP{}, and \IRE{} catalogs. Aside from the anticipated drop in the number of stars, note the shift favoring fainter stars in CELESTA due to the cuts of the giant stars that were more likely to be observed by the \emph{Hipparcos} mission.}
\label{MagsByCatalog}
\end{center}
\end{figure}

The magnitude distribution of CELESTA, the \IRE{}, and \RHIP{} can be seen in Figure \ref{MagsByCatalog}. Though CELESTA necessarily presents a smaller number of stars by way of the necessary cuts, we also see a shift in the focus of CELESTA towards the fainter end of the distribution due to the elimination of a greater number of brighter stars (the giants) that would have been easier for \emph{Hipparcos} to detect than the fainter stars. Conversely, there is bias against closer stars in the \IRE{} because the brightest stars are saturated in several input catalogs, resulting in insufficient data for modeling.

\begin{figure}
\begin{center}
\includegraphics[width=1.0\linewidth]{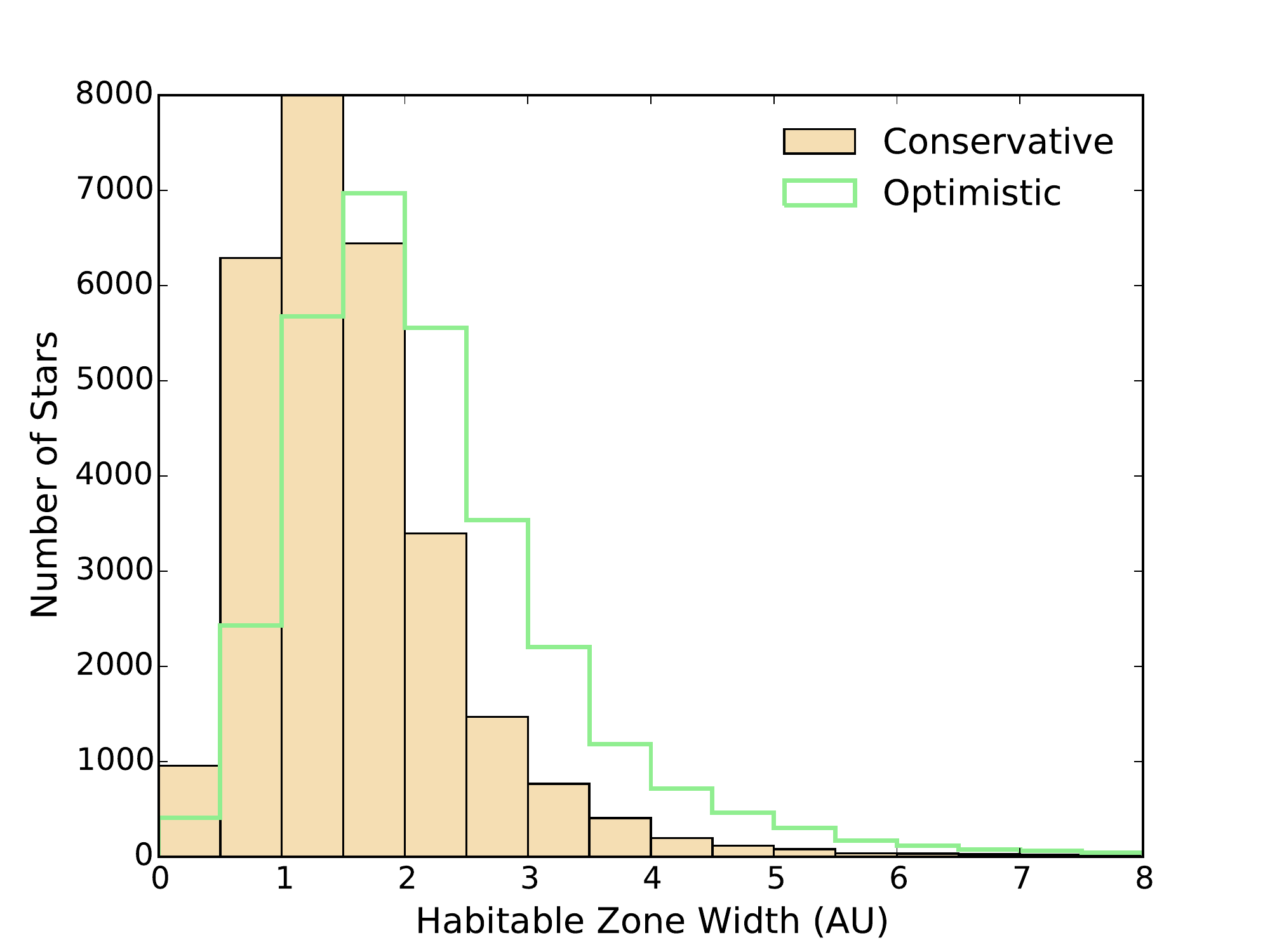}
\caption{The distribution of HZ widths found in CELESTA; note the conservative HZ widths dominate in the narrower end of the spectrum, whereas the optimistic zones are more widely distributed, favoring thicker HZs.}
\label{HZwidths}
\end{center}
\end{figure}

The HZ widths of Figure \ref{HZwidths} provide another example of perspective that can be gained from analysis of CELESTA. HZ widths tend to be less than 5 AU, with the majority falling between 1 AU \textendash{} 1.5 AU.

\begin{figure}
\begin{center}
\includegraphics[width=.9\linewidth]{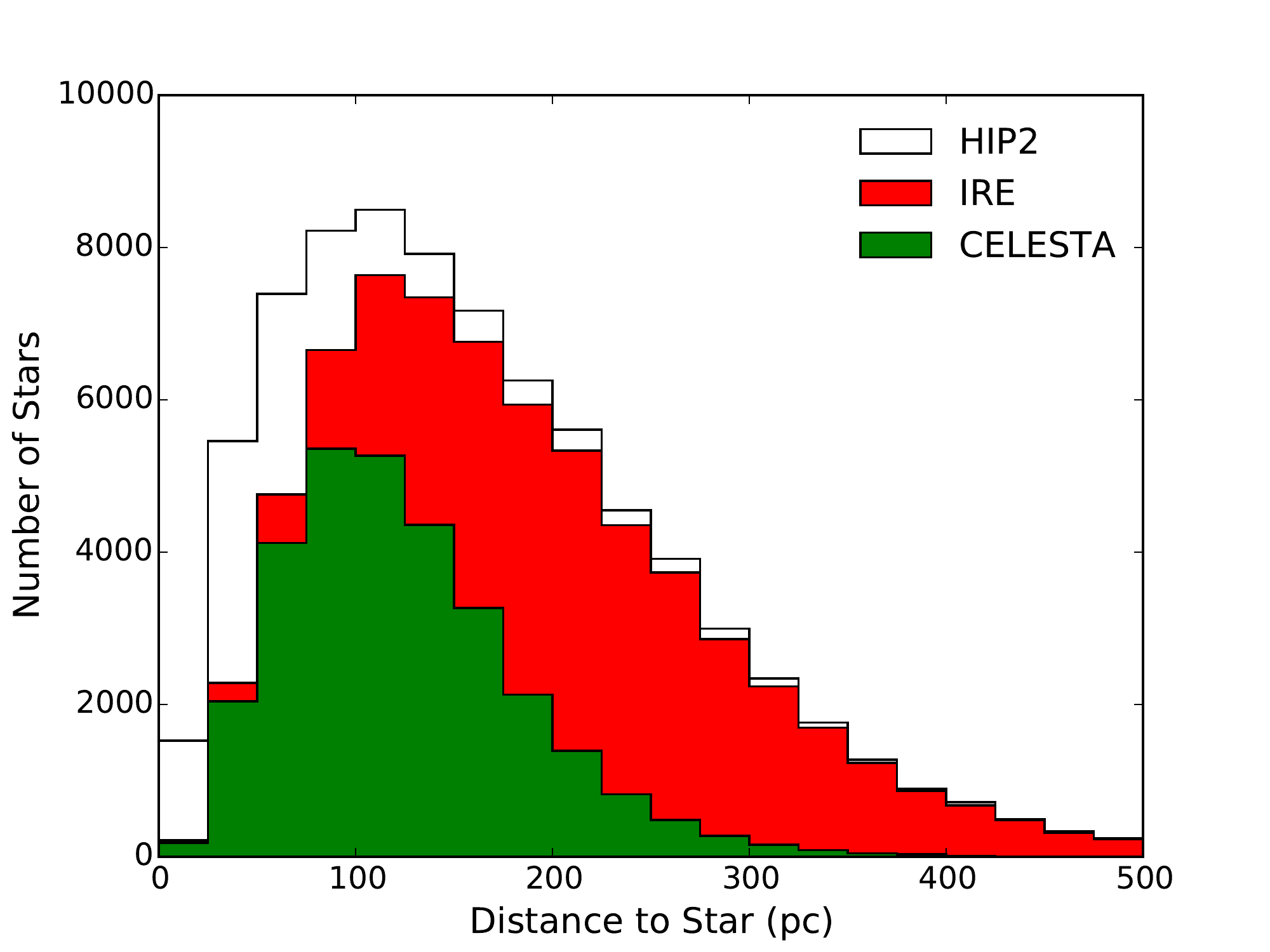}
\caption{Stellar distances found in CELESTA, the \IRE{}, and the \RHIP{}. The sensitivity of \emph{Hipparcos} to intrinsically fainter MS stars drops markedly after $\sim$100 pc. Some of the closest stars are lost in the \IRE{} as their photometry is heavily saturated in the catalogs we used.}
\label{StarsByDistance}
\end{center}
\end{figure}

Contrasting CELESTA stars with those of \RHIP{} and \IRE{}, we found that the CELESTA stars tended to be closer, as exhibited in Figure \ref{StarsByDistance}. This selection effect resulted from the MS requirement favoring stars close enough to reach the threshold magnitude of \emph{Hipparcos}.

\section{Limitations and Uncertainty Analysis}
\label{errors}
CELESTA is a starting point and will grow in both stellar quantity and accuracy of calculated parameters as more data becomes available from current and future surveys. Here we recap some of the limitations of CELESTA and introduce other considerations. As shown by \cite{Kane:2014ci}, the uncertainty in stellar parameters can have a non-negligible effect on the calculated locations of the Habitable Zones.

\subsection{Limitations}
\label{limitations}
As described in Section \ref{HIP2}, the stars contained in CELESTA are limited to those available in the \RHIP{}. As specified in Section \ref{starselection}, CELESTA contains MS stars; M\textendash{}dwarfs and giant stars are not contained in CELESTA at present. We showed in Section \ref{ParallaxError} that is is possible for stars to be erroneously included or excluded if their temperatures were within \TeffUncertainty{} of one of the cutoff temperatures. In section \ref{Applications} we will touch upon some considerations for different types of surveys, e.g. the necessity to consider the region of sky available to your survey.

Our model does not account for binary stars and so we excluded them (57 stars) from our analysis. We also did not consider the variability of a star in our calculations. Both features could be interesting to investigate in the future.

\subsection{Predetermined Uncertainty}
This section discusses sources of uncertainty that we relied upon but did not ourselves derive.
\label{ParallaxError}
The uncertainty in parallax was provided by the \emph{Hipparcos} catalog. As mentioned in Section \ref{starselection}, uncertainty in parallax was not used to disqualify any stars from CELESTA. However, for statistical work we chose to cap the error in parallax at 30\%, excluding over 70,000 stars. Improved parallax measurements will allow additional stars to be included; this will be discussed further in Section \ref{GAIA}. We utilized \EDE{} provided uncertainties in stellar temperature, stellar radius, and stellar mass when evaluation our predictions, and example of which is shown with Figure \ref{TeffTeff}.

\subsection{Ascertained Uncertainties}
\label{ascertained}

\begin{figure}
\begin{center}
\includegraphics[width=1\linewidth]{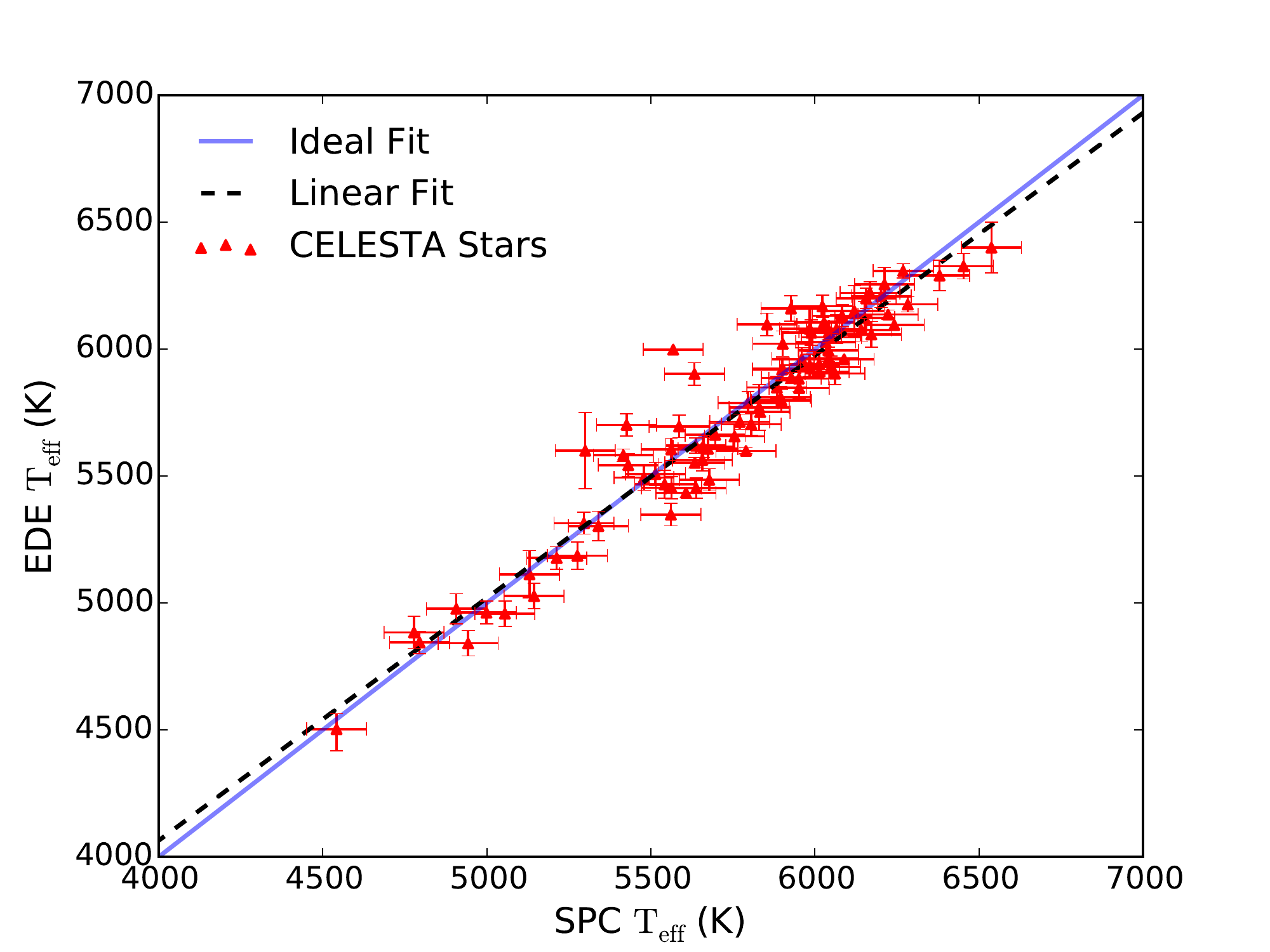}
\caption{CELESTA theoretical temperature predictions versus \EDE{} observational results. Ten outliers that were over $3 \sigma$ away from the linear fit (dashed line) were clipped. The solid blue line is the ideal fit, with a slope of $1/1$.}
\label{TeffTeff}
\end{center}
\end{figure}

We used a dataset exported on \exofiledate{} from The Exoplanet Data Explorer\footnote{http://www.exoplanets.org} service (\EDE; \citealt{Wright:2011vs}) that allowed us to validate our findings with \EDE{} supplied observational data, such as stellar temperatures and radii. For each planet, the downloaded file contained stellar temperature and uncertainty, a stellar binary flag, stellar radius and uncertainty therein, stellar distance and uncertainty therein, stellar mass, \emph{Hipparcos} number, and other parameters of convenience. Since \EDE{} is a catalog of planets, it was necessary to remove duplicate \emph{Hipparcos} stars for our analysis; the duplicate stars had identical stellar parameters.

Comparison of CELESTA and the \IRE{} with \EDE{} uncovered several trends. Of the 357 unique \emph{Hipparcos} stars in the \EDE{} dataset, 153 overlapped with the \IRE, revealing that over half of the \RHIP{} stars found in \EDE{} were not present in the \IRE. This is most likely due to the bias against closer (brighter) stars in the \IRE, as discussed in Section \ref{catalog}. CELESTA had 113 \emph{Hipparcos} numbers in common with the \EDE{} records, an expected consequence of cutting such a large number of stars from CELESTA. What follows are explanations of how we arrived at uncertainties for individual parameters.

Figure \ref{TeffTeff} provides a visual perspective on how we used standard deviation, based purely upon a linear fit between the \IRE{} and the \EDE{} effective temperatures, to determine the uncertainty in our calculated temperature to be \TeffUncertainty{}. To attain this uncertainty we recorded the absolute value of the difference between the predicted and observed temperatures for each star, and took the average of these differences to be our standard deviation $\sigma$:

\begin{equation}
\label{meanaverage}
\sigma =\frac{1}{N} \sum _{i=1}^N |x_i-y_i|.
\end{equation}

The initial deviation was $\pm 160$ K, but obvious outliers were present, and we chose $3\sigma$ as a clipping point to exclude the outliers. The following ten \emph{Hipparcos} stars were outside of the $3\sigma$ requirement: 8159 ($3.0\sigma$), 39417 ($3.8\sigma$), 40952 ($4.3\sigma$), 42723 $(3.4\sigma)$, 52409 ($3.8\sigma$), 54906 ($7.2\sigma$), 55664 ($4.3\sigma$), 61028 ($4.1\sigma$), 63584 ($6.5\sigma$), and 73146 $(4.3\sigma)$. This uncertainty tells us that some stars near the temperature boundaries could have been included or excluded in error. For example, 1,025 stars that were excluded and 1,129 stars that were included in CELESTA fell within the uncertainty of the 2,600 K \textendash{} 7,200 K limits. HZ and distance calculations also depend upon the calculated temperature and would be impacted as well, and these uncertainties are found in the complete catalog.

\begin{figure}
\begin{center}
\includegraphics[width=1\linewidth]{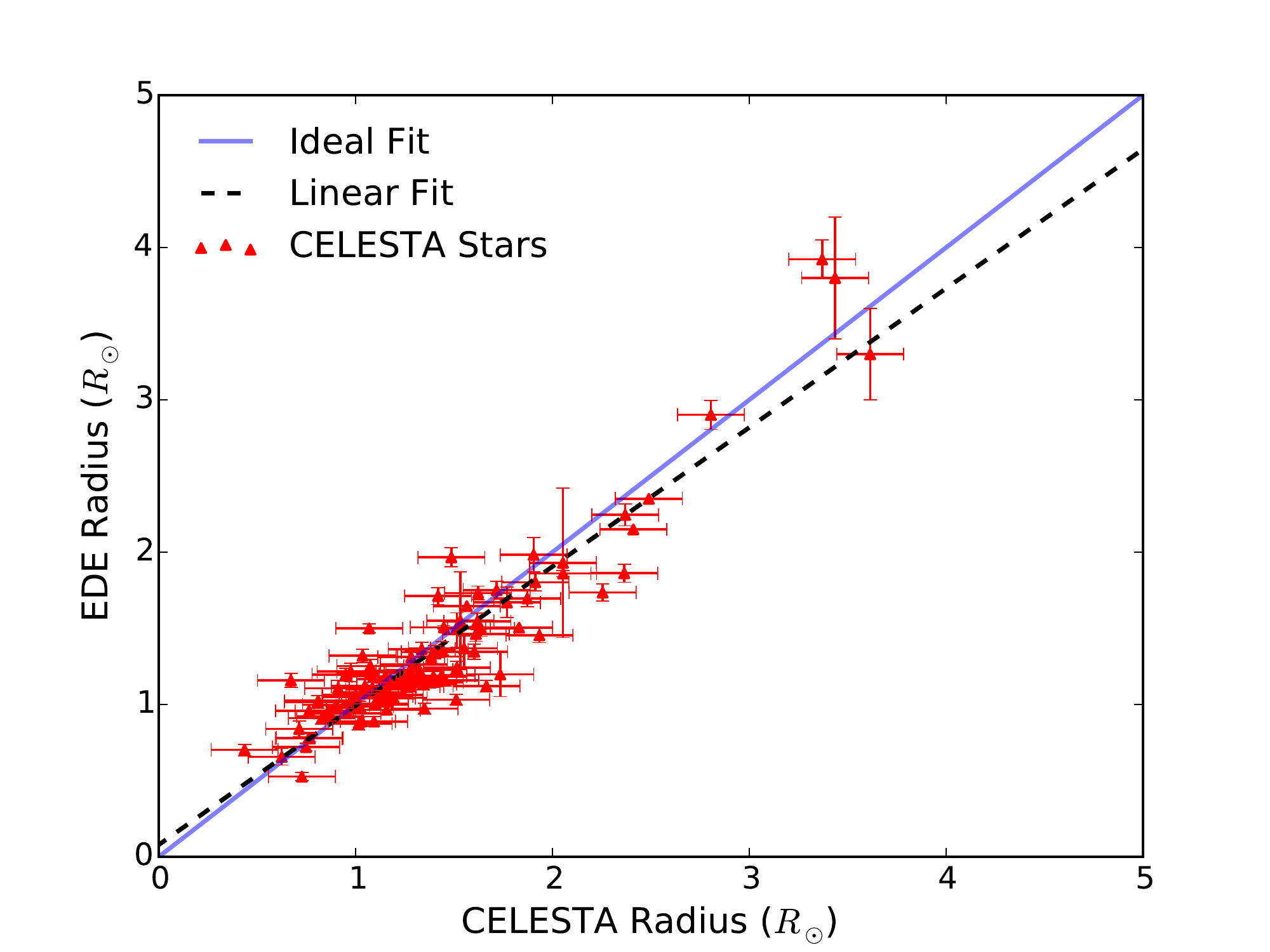}
\caption{CELESTA theoretical radius predictions versus \EDE{} observational results. Five outliers that were over $3 \sigma$ away from the linear fit (dashed line) were excluded. The solid blue line is the ideal fit, with a slope of $1/1$.}
\label{RadiusRadius}
\end{center}
\end{figure}

To ascertain the uncertainty in radius we applied the same method described above in determining the uncertainty in temperature, comparing the stellar radii calculated by CELESTA to those reported in the \EDE, as shown in Figure \ref{RadiusRadius}. The uncertainty in stellar radius was found to be \RadiusUncertainty. We removed five outliers that were over $3 \sigma$ away from the linear fit (dashed line) were excluded: 8159 $(3.2\sigma)$, 39417 $(3.2\sigma)$, 42723 $(3.6\sigma)$, 52409 $(3.5\sigma)$, and 63584 $(5.0\sigma)$.. 525 stars were excluded and 1,175 stars were included in CELESTA that lay within the radius uncertainty.

\begin{figure}
\begin{center}
\includegraphics[width=1\linewidth]{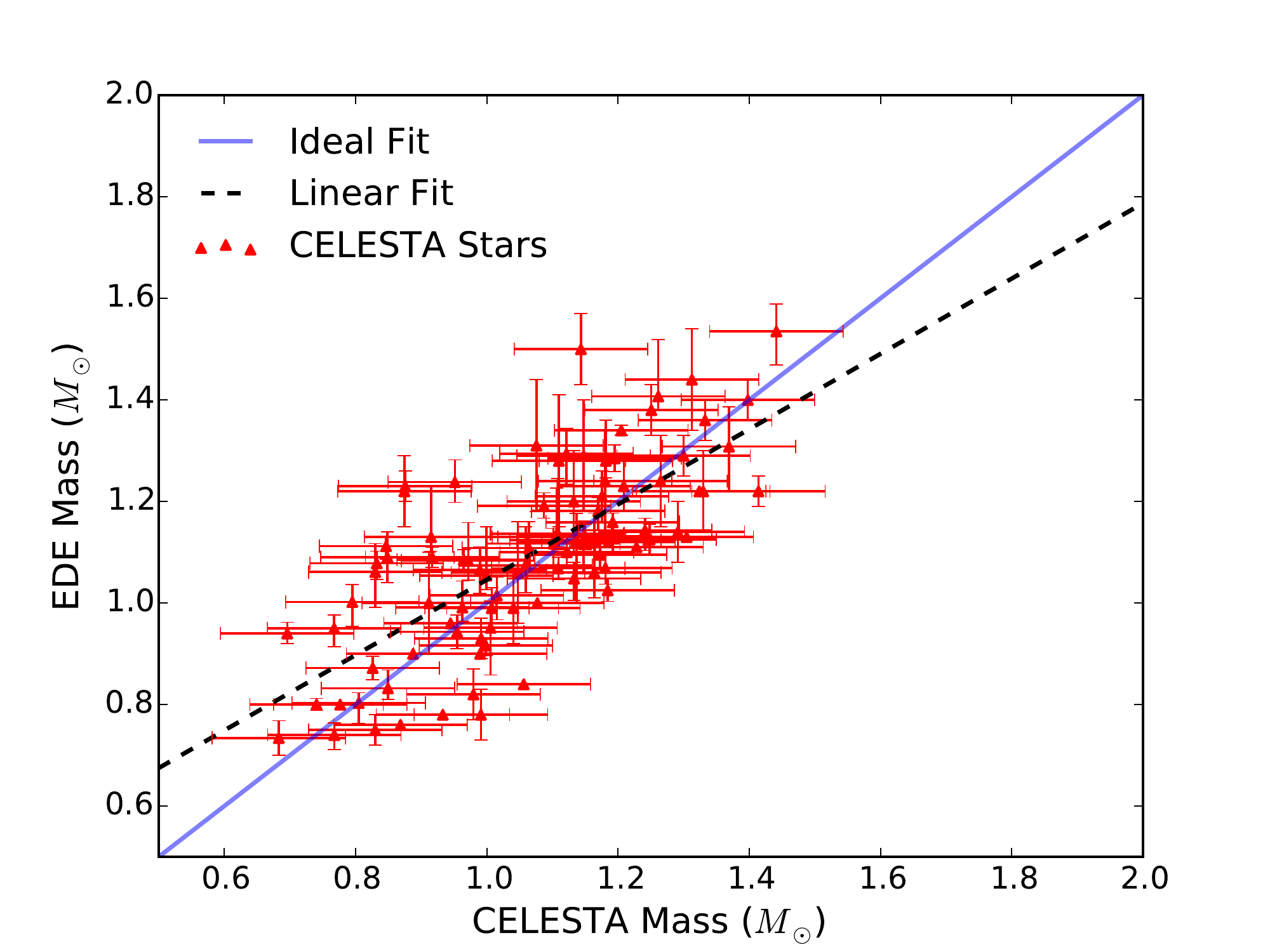}
\caption{CELESTA theoretical stellar mass predictions versus \EDE{} observational results. Eight outliers that were over $3 \sigma$ away from the linear fit (dashed line). The solid blue line is the ideal fit, with a slope of $1/1$.}
\label{MassMass}
\end{center}
\end{figure}

Similarly, as shown in Figure \ref{MassMass}, the uncertainty in stellar mass was determined to be $0.10 M_\odot$. Eight outliers past $3\sigma$ were excluded: 1640 $3.6\sigma)$, 42446 $(3.1\sigma)$, 43569 $(4.9\sigma)$, 54906 $(4.1\sigma)$, 63584 $(3.1\sigma)$, 80687 $(4.0\sigma)$, 84069 $(5.3\sigma)$, and 98714 $(6.1\sigma)$.

\subsection{Effective Reddening}
As in \citet{McDonald:2012tx}, interstellar reddening has not been accounted for due to the significant uncertainties in local 3D extinction maps. The magnitude of the correction for interstellar reddening depends on stellar color, hence stellar temperature. To test the effect of interstellar reddening, we have taken a typical CELESTA star (HIP1239; 6187 K) and de-reddened it using the approach adopted in \cite{McDonald:2009cu}. Additional reddening of $E(B-V) = 0.01$ mag increases the derived temperature by 50 K (0.8\%), the luminosity by 3.2\% and the radius by 1.7\%. The HZ distance will then increase by a similar factor.

There exists a strong similarity between the temperatures derived photometrically in CELESTA and (largely) spectroscopically in EDE (Figure \ref{TeffTeff}). Only a handful of stars scatter noticeably above the black dotted line (marking an exact correspondence). This shows that for most targets, interstellar reddening should be negligible compared to other uncertainties inherent in the data (e.g.\ parallax uncertainties). Confirmation of this can be found by comparing the 3D reddening maps from (e.g.) \citet{Lallement:2014fw}. A typical CELESTA star, at a distance of 125 pc, will only suffer $E(B-V) > 0.03$ mag in a few sightlines near the Galactic Plane. The effect of reddening becomes more strongly noticeable for stars beyond $\sim$300 pc \citep{McDonald:2012tx}, but for most stars in CELESTA other uncertainties provide a much larger contribution to the uncertainty in the HZ distance.

\subsection{Propogated Uncertainty}
\label{propogation}
To provide uncertainty in parameters dependent on uncertainties both provided and ascertained, we used the standard uncertainty propagation technique \citep{Taylor:1997td}
\begin{equation}
\label{propeq}
\delta q=\sqrt{\left(\frac{\partial q}{\partial x}\delta x\right)^2+\cdots +\left(\frac{\partial q}{\partial z}\delta z\right)^2}.
\end{equation}

Distance (in parsecs) was determined from parallax, \emph{p}
\begin{equation}
d=\frac{1}{p}.
\end{equation}

Propagation of uncertainty yielded the familiar 
\begin{equation}
\frac{\delta d}{d}=\frac{\delta p}{p}.
\end{equation}

Fractional uncertainty in luminosity is described by
\begin{equation}
\frac{\delta L}{L} = \sqrt{ \left(4 \frac{\delta T}{T}\right)^2 + \left(2 \frac{\delta d}{d}\right)^2 }
\end{equation}

HZ parameters were given by Equation \ref{seffequation}. Propagation of uncertainty led to
\begin{equation}
\delta S= \left(a+2 b T+3 c T^2+4 d T^3\right)\delta T.
\end{equation}

The HZ orbital distance uncertainty was calculated by propagating uncertainty through Equation \ref{distanceequation}, resulting in
\begin{equation}
\delta a_{\text{HZ}}=\frac{1}{2} \sqrt{\frac{L^2 \delta S_{\text{eff}}^2+\delta L^2 S_{\text{eff}}^2}{L S_{\text{eff}}^3 L_\odot}}
\end{equation}

The orbital period is given by Equation 5. Propagation of the uncertainty in $d$ and $M$ gives an uncertainty in $P$ of:
\begin{equation}
\delta P = (1.001\times 10^5) \sqrt{\frac{d^3 \delta M^2+d \delta a_{\text{HZ}}^2 M^2}{M^3}}.
\end{equation}

\subsection{Gaia: Future CELESTA Upgrades}
\label{GAIA}
In 2013 the European Space Agency launched the \emph{Gaia}\footnote{http://sci.esa.int/gaia/} satellite with a mission to map out the brightest billion stars \citep{deBruijne:2015wb}. Greatly increased precision parallax measurements ($\sim$ 24 $\mu$as) will start to be released by the \emph{Gaia} project initially in the form of of the \emph{Hundred Thousand Proper Motions} (HTPM) catalog, initially set to release 22 months after launch \citep{Eyer:2013uw}. Currently the ESA is estimating an initial release of data to take place in summer 2016\footnote{http://www.cosmos.esa.int/web/gaia/release}. As new parallax data becomes available, CELESTA calculations will become more precise. Note that as the catalog reaches further distances, interstellar reddening will need to be accounted for.


\section{Example Applications}
\label{Applications}

\begin{figure}[h]
\begin{center}
\includegraphics[width=1\linewidth]{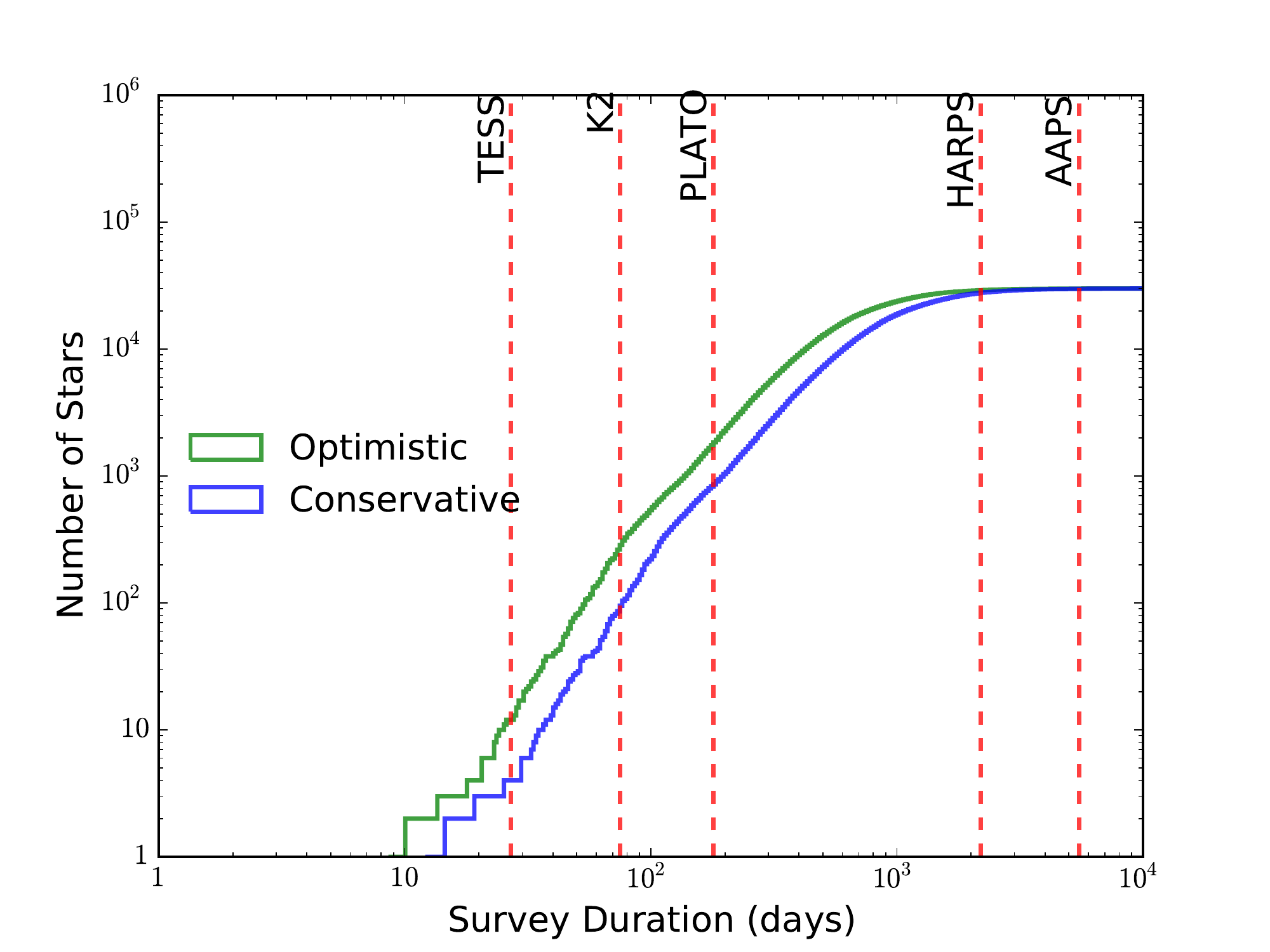}
\caption{
Plot showing prospective habitable zone penetration for a given survey duration. Surveys listed as reference points are TESS at 27 days, K2 at 75 days, PLATO at 180 days, HARPS at 6 years, and AAPS at 15 years. These are the maximum numbers of HZs probed (not the number of planets detected) assuming continuous observation of CELESTA stars by an all-sky survey.}
\label{SurveyROI}
\end{center}
\end{figure}

The primary purpose of CELESTA is as an input catalog for target selection with current and upcoming exoplanet surveys. For transit and radial velocity (RV) surveys, the sensitivity to planets within the OHZ and CHZ increase as the survey duration increases. Note that the sensitivity of a survey to HZ planets ceases to increase once the survey duration extends past the outer edge of the HZ. There are numerous surveys that are specifically targeting M dwarfs since such systems produce both larger RV and transit signatures for a planet of a given size/mass. In particular, the bias of these techniques towards shorter orbital periods lend themselves towards the discovery of HZ planets around cooler stars since the HZ is closer to the star in those cases. Ground\textendash{}based surveys of this type include the \emph{Habitable\textendash{}Zone Planet Finder} (HPF)  \citep{Mahadevan:2012dc}, the \emph{Miniature Exoplanet Radial Velocity Array} (MINERVA) \citep{Swift:2014vr}, and the \emph{MEarth} project \citep{Irwin:2008dz}.

Here we consider both ground and space\textendash{}based surveys that target a relatively broad range of spectral types, primarily for bright stars which dominate the \RHIP. Figure \ref{SurveyROI} plots the cumulative number of CELESTA stars for which a survey sensitivity lies within the OHZ (green line) and CHZ (blue line) as a function of the survey duration. The vertical dashed lines indicate the typical expected period sensitivity of several current and future surveys. Note that these vertical lines do not make any assumptions regarding the number of transits or RV measurements required for a particular orbital period to be considered a robust detection by that survey and so in some cases the sensitivity may be considerably less than that shown. These cumulative numbers also comprise an upper limit, assuming an all\textendash{}sky survey.

The TESS mission will observe each pointing for a total of 27 days \citep{Ricker:2015ie} which yields a total of 4 and 12 CELESTA stars with sensitivities within the CHZ and OHZ respectively. There will be overlap of TESS fields towards the celestial poles that is expected to result in greater period sensitivity in those regions. For example, TESS will continuously observe stars that lie within 17 degrees of the ecliptic pole, probing 101 HZs of the 590 CELESTA stars within that region. Note that the estimated number of stars for which the TESS mission will monitor HZs are slightly underestimated because TESS will observe a number of bright stars not currently included in CELESTA.

For the \emph{K2} mission, \cite{Howell:2014vu} anticipate campaigns lasting as long as 75 days. For such an observing duration, the sensitivity for the CHZ and OHZ stars is 87 and 267 respectively. The \emph{PLAnetary Transits and Oscillations of stars} (PLATO) mission \citep{Roxburgh:2007vxb} will have observing campaigns lasting $\sim$180 days. The campaigns will probe into the CHZ and OHZ for 828 and 1,747 stars respectively.

Finally, the ground\textendash{}based RV surveys by the HARPS \citep{Bonfils:2011dh} and AAPS \citep{Wittenmyer:2014tj} teams have now established an extensive time baseline making them sensitive to orbital periods that lie beyond the HZ for most stars. Figure \ref{SurveyROI} thus shows that the HZ sensitivity of these surveys for CELESTA stars lies where the CHZ and OHZ numbers converge to the maximum number of stars (with uncertainty in parallax under 30\%): $\sim$36,000.

The example applications described here assume that the surveys are limited to the bright stars contained within the \emph{Hipparcos} catalog. However, there are various surveys that monitor fainter stars, often to increase stellar density and planet yield for transit surveys. As noted in Section \ref{GAIA}, the \emph{Gaia} satellite will provide stellar information which will significantly expand the CELESTA catalog to much fainter magnitudes. The \emph{Gaia} upgrade will correspondingly enable CELESTA to be applicable to a broader range of exoplanet surveys and their expected yield of HZ planets.

\section{Conclusions}
\label{conclusions}
We have created the Catalog of Earth\textendash{}Like Exoplanet Survey TArgets (\catalogname), a repository containing stellar parameters and habitable zone orbital radii and periods for near\textendash{}Earth\textendash{}mass planets that may be found around \starquantity{} nearby stars. Comparison of theoretical calculations to stars with observational results showed an excellent fit in effective temperature (\TeffUncertainty{}). Analysis can be extended further as new data become available, e.g. upon the release of \emph{Gaia} parallax measurements, currently slated for release in summer 2016 \citep{Eyer:2013uw}. The final \emph{Gaia} data release of 2021 \citep{Eyer:2013uw} will coincide well with an upgraded CELESTA for the 2020 launch of \emph{Euclid}\footnote{http://sci.esa.int/euclid/} \citep{Penny:2013hg,McDonald:2014el}. Surveys can make immediate use of CELESTA for target selection and mission planning purposes by estimating the number of Habitable Zones probed for a given number of days committed for observation. For updated computations, survey scientists can return to the online edition of CELESTA, available online at a dedicated host\footnote{http://celesta.info} and at the VizieR repository\footnote{http://vizier.u-strasbg.fr}.


\section*{Acknowledgements}

The authors would like to thank the anonymous referee, Dr. Mark Jesus M Magbanua\footnote{University of California, San Francisco}, and Professor Eric Mamajek\footnote{University of Rochester} whose insights greatly improved the quality of the paper. Thank you to Dr. Christopher Burns\footnote{Carnegie Observatories, Pasadena} for allowing us to port his \emph{sigfigs.py} program. Thank you to Dr. Natalie Hinkel\footnote{Arizona State University, Tempe} for her insights. This research has made use of the following online resources: the Exoplanet Orbit Database \citep{Wright:2011vs} and the Exoplanet Data Explorer (http://exoplanets.org; \citealt{Han:2014hn}), the Habitable Zone Gallery (http://hzgallery.org; \citealt{Kane:2012co}), and the VizieR catalog access tool \citep{Ochsenbein:2000tjb}. The results reported herein benefited from collaborations and/or information exchange within NASA's Nexus for Exoplanet System Science (NExSS) research coordination network sponsored by NASA's Science Mission Directorate.

\bibliography{Papers.bib}

\end{document}